%% file: main.tex
\documentclass[conference,10pt]{IEEEtran}

\usepackage[subrefformat=parens,labelformat=parens,caption=false,font=footnotesize]{subfig}
\usepackage{array}
\usepackage{amsmath}
\usepackage{breqn}
\usepackage[nolist,nohyperlinks]{acronym}
\usepackage{graphicx,wrapfig,lipsum}
\usepackage{rotating}
\usepackage{cleveref}
\usepackage{algpseudocode}
\usepackage{epstopdf}
\usepackage{balance}
\usepackage{amsmath, amssymb, amsthm}
\usepackage{stmaryrd}
\usepackage{verbatim}
\usepackage[short]{optidef}
\usepackage{url}
\usepackage[utf8]{inputenc}
\usepackage[english]{babel}

\usepackage{multicol}
\usepackage{xr-hyper}

\usepackage{scalerel}
\usepackage{multicol}

\usepackage[noadjust]{cite}
\usepackage[normalem]{ulem}
\usepackage{color}
\usepackage[linesnumbered,ruled,vlined]{algorithm2e}
\usepackage{dsfont}
\usepackage{bm}
\usepackage[short]{optidef}
\usepackage{diagbox}
\usepackage{enumitem}
\usepackage{slashbox}
\usepackage[ruled]{algorithm2e}
\usepackage{multirow}
\usepackage[T1]{}
\usepackage{color, colortbl}
\usepackage{babel}
\usepackage{verbatim}
\usepackage{textcomp}
\usepackage{tabulary}
\usepackage{caption}
\usepackage{bbm}
\definecolor{Gray}{gray}{0.95}
\definecolor{LightCyan}{rgb}{0.8,0.85,1}
\definecolor{LightBlue}{rgb}{0.6,0.6,1}
\usepackage[font=small]{caption}

\setlist{nosep}

\usepackage{mathtools}

\begin{document}
\title{A Multi-Armed Bandit Framework for Online Optimisation in Green Integrated Terrestrial and Non-Terrestrial Networks}

\author{Henri Alam$^{\dag\ddag}$, Antonio De Domenico$^\dag$, Tareq Si Salem$^\dag$, Florian Kaltenberger$^\ddag$\\
 \small{$^\dag$Huawei Technologies, Paris Research Center, 20 quai du Point du Jour, Boulogne Billancourt, France.} \\
 \small{$^\ddag$EURECOM, 2229 route des Crêtes, 06904 Sophia Antipolis Cedex, France.} \\
 }
\maketitle

\thispagestyle{empty}

\begin{abstract}

Integrated \ac{TN-NTN} architectures offer a promising solution for expanding coverage and improving capacity for the network. While \acp{NTN} are primarily exploited for these specific reasons, their role in alleviating \ac{TN} load and enabling energy-efficient operation has received comparatively less attention. In light of growing concerns associated with the densification of terrestrial deployments, this work aims to explore the potential of \acp{NTN} in supporting a more sustainable network. In this paper, we propose a novel online optimisation framework for integrated \ac{TN-NTN} architectures, built on a \ac{MAB} formulation and leveraging the \ac{BCOMD} algorithm. Our approach adaptively optimises key system parameters—including bandwidth allocation, \ac{UE} association, and \ac{MBS} shutdown—to balance network capacity and energy efficiency in real time. Extensive system-level simulations over a $24$-hour period show that our framework significantly reduces the proportion of unsatisfied \acp{UE} during peak hours and achieves up to $19~\%$ throughput gains and $5~\%$ energy savings in low-traffic periods, outperforming standard network settings following \acs{3GPP} recommendations.

\end{abstract}

\section{Introduction}
\label{sec:intro}
Recent advancements in cellular communications have sharply increased the demand for high-speed data, driving the need for broader network coverage and higher capacity. To address these challenges, mobile operators have intensified their deployment of terrestrial \acp{MBS}. 
This constant expansion has resulted in increased energy consumption, raising environmental concerns from a societal perspective as well as economic challenges for network operators.
Thus, minimising energy consumption while maintaining \ac{QoS} standards has become a key objective in mobile network management \cite{Lopez_Perez_2022_Survey}. \newline
\Acp{NTN} have emerged as a practical solution to complement \acp{TN} and extend coverage to underserved areas in the past few years. \acp{NTN} use airborne platforms such as drones or satellites as relay nodes or \acp{MBS} to provide connectivity to user equipment \acp{UE}. Their key advantage lies in offering wide-area coverage, particularly in remote regions where deploying terrestrial \acp{MBS} is costly or impractical. In particular, \ac{LEO} satellites are poised to play a central role in delivering high-capacity space-based connectivity \cite{Giordani_2021}, as they benefit from reduced latency, stronger signals, and lower energy requirements for both launch and communication, effectively building an integrated \ac{TN-NTN} capable of delivering seamless, high-capacity communication services while ensuring efficient and reliable connectivity for the \acp{UE} \cite{Benzaghta_2022}. \newline
Typically, these \acp{UE} are associated with the \ac{MBS} offering the highest \ac{RSRP}. However, this approach ignores traffic demand variations, often leading to poor load distribution and degraded network performance. An effective policy for load balancing usually involves a pricing-based association strategy \cite{Shen_2014} which considers both signal quality and cell load.
Alam et al.~\cite{Alam_2023} propose a similar approach in the context of an integrated \ac{TN-NTN}, leveraging satellite resources to distribute the load to improve overall network capacity and coverage. \newline 
Moreover, keeping all \acp{MBS} active during low traffic can lead to inefficient energy and resource use, as many may be underutilised. In an integrated \ac{TN-NTN}, selectively turning off some \acp{MBS} and offloading \acp{UE} to satellites can help reduce energy consumption. To that end, our previous work \cite{Alam_2025} proposed a framework designed to balance network fairness and energy
consumption by adjusting to varying traffic conditions in an integrated \ac{TN-NTN}.\newline
In this paper, we propose a novel framework that dynamically balances network capacity and terrestrial energy consumption in an integrated \ac{TN-NTN} architecture. The framework is formulated as a \ac{MAB} problem and leverages a constrained online learning algorithm named \ac{BCOMD}, introduced in \cite{Si_Salem_2025}, to adaptively select the optimal configuration of system parameters in response to time-varying traffic demands. By jointly optimising UE association, bandwidth allocation, and \ac{MBS} shutdown decisions, the framework effectively improves load distribution, reduces energy usage during low-traffic periods, and enhances user satisfaction in high-traffic scenarios.

\section{System Model}
We consider a \ac{DL} cellular network, operating over $T$ time slots, which consists of $M$ terrestrial \acp{MBS} and $N$ \ac{LEO} satellites mounted with \acp{MBS}, for a total of $L$ MBSs. $K$ \acp{UE} are deployed in the study area. The network operates in the S-band at approximately $2$ GHz, where the total available system bandwidth $W$ is allocated by the mobile network operator between terrestrial and non-terrestrial tiers, each using orthogonal frequency bands. Throughout this paper, we denote the set of terrestrial \acp{MBS} by $\mathcal{T}$ and the set of satellite \acp{MBS} by $\mathcal{S}$. The set of UEs is represented as $\mathcal{U} = \{1, \dots, K\}$, and the complete set of \acp{MBS} is given by $\mathcal{B} = \mathcal{T} \cup \mathcal{S} = \{1,\dots, L\}$. For the channel model, the large-scale channel gain between a terrestrial \ac{MBS} $j$ and a UE $i$ is computed as follows:
\begin{equation}\label{channel_terrestrial}
\begin{split}
\beta_{ij} &= \mathcal{M}\left( G_{\mathrm{T_X}} + G_{\mathrm{UE}} + \mathrm{PL}_{ij}^b + \mathrm{SF}_{ij} + \mathrm{PL}_{ij}^{\mathrm{tw}} + \mathrm{PL}_{ij}^{\mathrm{in}} \right. \\
     &\quad \left. + \mathcal{N}\left( 0, \sigma_p^2 \right) \right).
\end{split}
\end{equation}
where all terms are in dB and are mapped to linear space using the operator $\mathcal{M}(\cdot)$. Here, $G_{\mathrm{UE}}$ and $G_{\mathrm{T_X}}$ represent the \ac{UE} and \ac{MBS} antenna gain respectively, $\mathrm{PL}_{ij}^b$ is the basic outdoor path loss \cite[Table 7.4.1-1]{3GPPTR38.901}, and $\mathrm{SF}_{ij}$ is the shadow fading, normally distributed with mean $0$ and variance $\sigma_{\mathrm{SF}}^2$. The remaining terms ($\mathrm{PL}_{ij}^{\mathrm{tw}}$, $\mathrm{PL}_{ij}^{\mathrm{in}}$, and $\mathcal{N}(0,\sigma^2_p)$) account for building penetration losses, detailed in \cite{3GPPTR38.901}. The \ac{LoS} condition for each \ac{UE} is computed as in \cite[Table 7.4.2-1]{3GPPTR38.901}.
In contrast, for a satellite \ac{MBS} $j$ serving \ac{UE} $i$, the large-scale channel gain can be expressed as follows \cite{3GPPTR38.811}:
\begin{equation}\label{channel_satellite}
     \beta_{ij}  = \mathcal{M} \left( G_{\mathrm{T_X}} + G_{\mathrm{UE}} + \mathrm{PL}_{ij}^b + \mathrm{SF}_{ij} + \mathrm{CL} + \mathrm{PL}_{ij}^s + \mathrm{PL}_{ij}^e \right)
\end{equation}
In \eqref{channel_satellite}, $\mathrm{CL}$ denotes clutter loss, which is the attenuation caused by buildings and vegetation near the \ac{UE}, and $\mathrm{PL}_s$ represents scintillation loss, capturing rapid fluctuations in signal amplitude and phase due to ionospheric conditions. Lastly, $\mathrm{PL}_{ij}^e$ denotes the building entry loss, representing the attenuation experienced by all \acp{UE} positioned indoors.
Since interference between both terrestrial and non-terrestrial tiers is negligible due to orthogonal bandwidth allocations, the large-scale \ac{SINR} for each \ac{UE} $i$ can be calculated as follows:
\begin{equation}\label{SINR}
    \gamma_{ij}  = \frac{ \beta_{ij} p_j}{ \sum\limits_{\substack{ j^' \in \mathcal{I}_j}} \beta_{ij^'} p_{j^'} + \sigma^2},
\end{equation}
where $p_j$ represents the transmit power per \ac{RE} allocated by \ac{MBS} $j$, $\mathcal{I}_j$ denotes the set of \acp{MBS} interfering with serving \ac{MBS} $j$, and $\sigma^2$ accounts for the noise power per \ac{RE}.
In our study, each \ac{UE} $i$ has a specific data-rate demand $\rho_i$, modelled as a random variable following an exponential distribution of parameter $\lambda_{\mathrm{U}}$.
Then, the number of \acp{PRB} assigned to the \ac{UE} by the associated \ac{MBS} $j$ is computed as:
\begin{equation}\label{PRB_allocation}
   B_{ij}  =   \left\lceil   \frac{\rho_i}{\Delta \log_2\left(1 + \gamma_{ij}\right)} \right\rceil.
\end{equation}
In \eqref{PRB_allocation}, the denominator is the product between the spectral efficiency and $\Delta$, which represents the total bandwidth of a single \ac{PRB} in 5G \ac{NR}. Finally, $\left\lceil \cdot \right\rceil$ denotes the ceiling function, which rounds up the input number to the nearest integer. The load $\nu_j$ for \ac{MBS} $j$ is then defined as the fraction of \acp{PRB} being utilised.
Using this, we can calculate the mean throughput for \ac{UE} $i$ served by \ac{MBS} $j$ as:
\begin{equation}\label{Shannon_data_rate_PRB}
    R_{ij}  = \Delta B_{ij} \log_2(1 + \gamma_{ij}).
\end{equation}
Finally, the terrestrial \ac{MBS} energy consumption model, dependent on various parameters as described in \cite{Piovesan2022}, can be succinctly represented as the sum of three components. 
The baseline component refers to the energy consumed by elements that remain active even when the \ac{MBS} is shut down. The static component is the fixed energy consumption required to maintain essential systems operational, independent of traffic load. Lastly, the dynamic component varies with the traffic load, increasing when the \ac{MBS} transmits at higher power levels or utilises additional \acp{PRB}.
For a \ac{MBS} $j$, this model is written as:
\begin{equation}\label{Power_conso_model_v1}
    Q_j = P_0 + p_j + \psi_j \mathbbm{1}_{\{p_j>0\}},
\end{equation}
where $P_0$ denotes the baseline energy consumption, $\psi_j$ indicates the static component, and $p_j$, the transmission power of \ac{MBS} $j$, corresponds to the dynamic component. Additionally, $\mathbbm{1}_{\{\cdot\}}$ is the indicator function that equals $1$ if the inputted condition is True, and $0$ otherwise. 
The satellite is presumed to harvest its energy from solar panels.
We also define the parameters that we intend to optimise in our framework: $\varepsilon$ is the proportion of the bandwidth which is allocated to the \ac{LEO} satellites. $\tau_\nu$ is the threshold considered for the load of a \ac{MBS}, which determines whether we attempt to shut it down or not, while $\tau_{\mathrm{RSRP}}$ is also a threshold for the perceived \ac{RSRP}. Finally, $\alpha$ is a weight that controls the influence of \ac{MBS} load on the \ac{UE} association decision. The role of each parameter will be explained in more detail in Section~\ref{sec:UtilityOptimization}.

\section{Problem Formulation}
Similar to the model proposed by Alam et al.~\cite{Alam_2025}, we aim to design a framework that jointly optimises network capacity and \ac{TN} energy consumption by dynamically adjusting resource allocation based on network load, while satisfying the data-rate requirements of each \ac{UE}. \newline
More specifically, our objective is to identify the optimal policy, defined as an action sequence selected from a set of $n$ distinct configurations of $\theta = \left[\varepsilon,\tau_\nu,\tau_{\mathrm{RSRP}},\alpha\right]$, each referred to as an \textit{arm}. Indeed, each arm represents a specific setting for those parameters, chosen from the $n$ different combinations possible. 
These arms directly impact network behaviour through a heuristic, which is detailed in Section~\ref{sec:UtilityOptimization}.
To evaluate the performance of the network at time $t$, we define a cost function that captures the key trade-offs introduced by selecting a given arm $a_t$, which is drawn according to a probability distribution $x_t$ over the action space $\Delta_n$ (the $n$-dimensional probability simplex):
\begin{equation}\label{reward_function}
    f_t\left( a_t , x_t\right) = \zeta \sum_{j\in\mathcal{B}} Q_j(\theta) - \sum\limits_{i \in \mathcal{U}} \log(R_i(\theta)),
\end{equation}
where $R_i(\theta)$ denotes the throughput perceived by \ac{UE} $i$ and $Q_j(\theta)$ represents the energy consumption of \ac{MBS} $j$, both for a given configuration $\theta$.
$\zeta$ is a regularisation factor that allows balancing the trade-off between \ac{UE} performance (i.e., \ac{SLT}) and network energy consumption.
Without loss of generality, the cost function values are normalised between $0$ and $1$.
In parallel, we define the constraint violation incurred by selecting arm $a_t$ as:
\begin{equation}\label{violation_function}
    g_t\left( a_t, x_t\right) = \frac{1}{K} \sum\limits_{i \in \mathcal{U}} \mathbbm{1}_{\{ R_i < \rho_i \}}.
\end{equation}
Note that if a \ac{UE} perceives a \ac{RSRP} lower than a set threshold $\mathrm{RSRP}_{\rm min}$, it is considered unsatisfied. \newline
Naturally, the cost distribution associated with each arm evolves over time as network demand fluctuates with the traffic load. Indeed, an arm that enhances capacity under high traffic may be suboptimal in low-traffic scenarios—emphasising the need for context-aware arm selection.
This non-stationarity in cost aligns well with the adversarial bandit-feedback setting proposed in \cite{Si_Salem_2025}. Accordingly, we leverage the algorithm introduced in \cite{Si_Salem_2025} to handle dynamic costs while ensuring long-term constraint satisfaction. \newline
We denote by $\left\{x^*_t\right\}^T_{t=1}$ the \textit{oracle} policy, i.e., the action sequence that achieves the minimum cumulative loss over the time horizon:
\begin{equation}\label{oracle_policy}
\left\{ x_t^\star \right\}_{t=1}^T \in 
\arg \min_{\left\{ x_t \right\}_{t=1}^T  \in \bigtimes_{t=1}^{T} \Delta_{n,t}} \ 
\left\{ \sum_{t=1}^T f_t(a_t, x_t) \right\},
\end{equation}
where $\Delta_{n,t}$ is the set of feasible points within the simplex at time $t$:
\begin{equation}\label{feasible_set}
 \Delta_{n,t} \triangleq \left\{ x \in \Delta_n : g_t\left(a_t,x \right) = 0 \right\}.
\end{equation}

Our goal is to find a policy $\pi$ that minimises cumulative loss relative to the oracle, while also satisfying time-varying constraints. Adopting the notation from \cite{Si_Salem_2025}, we define the regret and constraint violation which we want to minimise as:
\begin{equation}\label{regrets_cost}
\mathfrak{R}_T(\pi) \triangleq \mathbb{E}_{\pi} \left[ \sum_{t=1}^T f_t(a_t, \pi_t ) \right] 
- \sum_{t=1}^T f_t(a_t, x^*_t),
\end{equation}
\begin{equation}\label{regrets_violation}
\mathfrak{V}_T(\pi) \triangleq \mathbb{E}_{\pi} \left[ \sum_{t=1}^T g_t(a_t, \pi_t) \right]. 
\end{equation}



\section{Finding Optimal Policy Using BCOMD}
\label{sec:UtilityOptimization}

In this section, we present the designed framework, which takes as input the set of parameters $\theta$ and associates a cost that quantifies both the network performance and the quality of the parameter configuration. Then, we describe the method used to determine the optimal setting of $\theta$, which minimises Eq. \eqref{regrets_cost} and \eqref{regrets_violation}.

\subsection{Network Optimisation Framework}\label{subseq:framework}

The framework proposed to measure network performance given $\theta$ can be broken down into the following steps:
\begin{enumerate}
    \item \textbf{Initialisation:} Given the input $\theta = \left[\varepsilon,\tau_\nu,\tau_{\mathrm{RSRP}},\alpha\right]$, we first associate each \ac{UE} using the max-\ac{RSRP} criterion, compute the resulting load on each \ac{MBS}, and redistribute the resources based on $\varepsilon$.
    \item \textbf{\ac{UE} association:} For each \ac{UE} $i$, we propose a pricing function which takes into account the load of \ac{MBS} $j$:
    \begin{equation}\label{Pricing_function}
    P_i(j) = \mathrm{RSRP}_{ij} - \alpha \nu_j.
    \end{equation}
    A positive value of $\alpha$ discourages highly loaded \acp{MBS} from serving additional \acp{UE}, thereby promoting load balancing across the terrestrial network. Conversely, a negative value encourages \acp{UE} to associate with loaded \acp{MBS}, leading to a higher number of inactive \acp{MBS}.
    We associate the \ac{UE} to the \ac{MBS} which maximises the pricing function.
    \item \textbf{\ac{MBS} Shutdown:} For each terrestrial \ac{MBS} $j$, we check if the sum of its load and the load of the satellite is smaller than $\tau_\nu$. If true, and all \acp{UE} served by this \ac{MBS} perceive a \ac{RSRP} greater than $\tau_{\mathrm{RSRP}}$ from the satellite, we handover the \acp{UE} to the satellite and shutdown the \ac{MBS}.
    \item \textbf{Cost and Constraint:} We compute the incurred cost and constraint violations based on \eqref{reward_function} and \eqref{violation_function}.
\end{enumerate}
\subsection{Bandit-feedback Constrained Online Mirror Descent}\label{subseq:OMD_MAB}
We now give an overview of the \ac{BCOMD} algorithm originally presented in \cite{Si_Salem_2025}, which works closely with the framework presented in Section \ref{subseq:framework} to derive the optimal policy.\newline
Firstly, by denoting $f_t \in [0,1]^n$ and $g_t \in [0,1]^n$ as the cost and constraint violation vectors, we can compute the expected loss and constraint violation at time $t$ respectively as:
\begin{equation}\label{expected_cost_violation}
f_t(x) \triangleq f_t \cdot x, \quad g_t(x) \triangleq g_t \cdot x,
\end{equation}
where $x \in \Delta_n$.\newline
The algorithm leverages a Lagrangian function defined as:
\begin{equation}\label{Lagrangian_func}
     \Psi(x,\lambda) \triangleq f_t(x) + \lambda g_t(x).
\end{equation}
The first term of the Lagrangian function corresponds to the cost function, while the second term imposes a penalty for soft constraint violations using the Lagrange multiplier, which acts as a weighting factor.
In the bandit-feedback setting, we derive unbiased estimators for the gradients of $f_t(x)$ and $g_t(x)$ as follows:
\begin{equation}
    \tilde{f}_t = \frac{f_t\left( a_t , x_t\right)}{x_{t,a_t}} e_{a_t}, \quad 
    \tilde{g}_t = \frac{g_t\left( a_t , x_t\right)}{x_{t,a_t}} e_{a_t},
\end{equation}
where $a_t$ is the arm selected at time step $t$, and $e_{a_t}$ denotes the unit basis vector corresponding to that arm.
However, the unbounded variance of these estimators poses significant challenges in establishing reliable performance guarantees under the bandit regime. To address that, \cite{Si_Salem_2025} proposes to use OMD, as it has proven to be effective in controlling the variance while claiming enhanced convergence speed compared to classic online gradient methods.\newline
In OMD, the updates are first performed in the dual space and then projected back to the primal space via a mirror map (e.g. the negative entropy).
Algorithm \ref{alg:bcomd} outlines the iterative refining of the action distribution through an OMD framework. Indeed, at each iteration, the update direction is determined by combining a gradient estimate of the cost function with a weighted estimate of the constraint gradients (Lines $21$-$22$). These weights are not fixed and are adaptively modified based on the accumulated constraint violations (Line $23$). This dynamic adjustment allows the policy to effectively balance cost minimisation with constraint satisfaction over time. Note that the probability of selecting any action is maintained above a predefined threshold $\gamma$.
The \ac{BCOMD} algorithm, adapted to our framework, is outlined in Algorithm \ref{alg:bcomd}.
\begin{algorithm}
\footnotesize
\caption{BCOMD - Network Optimisation}
\label{alg:bcomd}
\KwData{Initial $x_1 = (1/n)_{a \in \mathcal{A}}$, $\lambda_1 = 0$, Mirror map $\Phi : \mathbb{R}^n \to \mathbb{R}$, learning rate $\eta > 0$, $\gamma \in [0, 1/n]$, $\Omega > 0$}
\For{$t = 1,\dots,T$}{
    \textbf{Sample} action $a_t \sim x_t$ \tcp{Run Framework}
    \KwData{K UEs, L MBs, $\theta = \left[\varepsilon,\tau_\nu,\tau_{\mathrm{RSRP}},\alpha\right]$.}
    \textbf{Initialisation:}
    Association done through max-RSRP\;
    Compute the load for \ac{MBS}\; 
    Redistribute the resources according to $\varepsilon$\;
    \textbf{\ac{UE} association:}
        \For{all \acp{UE} $u$}{
            Associate \ac{UE} $u$ to \ac{MBS} $j^*$ such that: \\
            $j^* = \arg\max_j  P_u(j) $  \hspace*{2cm} \eqref{Pricing_function}
        }
        Recompute the load for \ac{MBS}\; 
        \textbf{\ac{MBS} Shutdown:}
        
        \For{all \acp{MBS} $j$}{
            \If{ $ \nu_j + \nu_{\mathrm{sat}}  \leq \tau_\nu$}{
                    \If{RSRP from satellite of all served \acp{UE} by \ac{MBS} $j$ $\geq \tau_{\mathrm{RSRP}}$ }{
                        Associate each \ac{UE} to the satellite\;
                        Shutdown \ac{MBS} $j$\;
                    }
            }
        }
    \textbf{Incur} $f_t(a_t,x_t)$ and $g_t(a_t,x_t)$ \tcp*{Bandit-feedback}
    
    $\tilde{f}_t \leftarrow \left( f_t(a_t,x_t)/x_{t,a_t} \right) e_{a_t}$ \tcp*{Loss gradient estimate}
    
    $\tilde{g}_t \leftarrow \left( g_t(a_t,x_t)/x_{t,a_t} \right) e_{a_t}$ \tcp*{Constraint gradient estimate}
    
    $\tilde{\omega}_t \leftarrow (\Omega / x_{t,a_t}) e_{a_t}$ \tcp*{Bias term}
    
    $\tilde{b}_t \leftarrow \tilde{\omega}_t + \tilde{f}_t + \lambda_t \tilde{g}_t$ \tcp*{Gradient for $\Psi(\cdot, \lambda_t)$}
    
    $y_{t+1} \leftarrow (\nabla \Phi)^{-1} \left( \nabla \Phi(x_t) - \eta \tilde{b}_t \right)$ \tcp*{Update primal action distribution}
    
    $x_{t+1} \leftarrow \Pi_{\Delta_{n,\gamma}}(y_{t+1})$ \tcp*{Project to feasible simplex}
    
    $\lambda_{t+1} \leftarrow \left( \lambda_t + \mu g_t(a_t) \right)_+$ \tcp*{Update dual variable}
}
\end{algorithm}

\section{Simulation results and analysis}\label{sec:Simulation_Results}
In this section, we assess the performance of our framework over $24$ hours, with the number of \acp{UE} varying at each hour, similar to \cite{Alam_2025}.
Using a custom-built system-level simulator, we collected $7\cdot10^3$ snapshots of the network for each hour of the day, yielding a total of $168\cdot10^3$ samples.
Then, we used the learned policy to sample an action for each hour of the day to evaluate the resulting performance.
$\varepsilon$ and $\tau_\nu$ take values in $\left[ 0.25, ~0.50, ~0.75, ~0.85, ~0.90 \right]$, while $\tau_{\mathrm{RSRP}}$ and $\alpha$ take values in $\left[ -80, ~-90, ~-100, ~-110, ~-120 \right]$ and $\left[ -3, ~-2, ~-1, ~0, ~1, ~2, ~3   \right]$ respectively.
Our study focuses on an area of approximately $2500$ $\mathrm{km}^2$, corresponding to the coverage area of an LEO satellite beam \cite{3GPPTR38.821}, and encompasses both urban and rural regions.
Additionally, we assume that the LEO constellation employs Earth-fixed beams \cite{3GPPTR38.811}.
The \acp{UE} are deployed uniformly across the study area, with a higher density in the urban region compared to the rural area. Similarly, the terrestrial \acp{MBS} are arranged in a hexagonal grid layout in both urban and rural areas, with a higher density of \acp{MBS} in the urban area.
Two benchmark configurations are provided to compare performances: the \texttt{3GPP-TN} scenario, where no satellite tier is present and the terrestrial tier is allocated a total bandwidth of $10$ MHz, and the \texttt{3GPP-NTN} scenario, where the total bandwidth is allocated as per \acs{3GPP} specifications \cite{3GPPTR38.821}, with $30$ MHz for the satellite tier and $10$ MHz for the terrestrial tier. In both scenarios, each \ac{UE} associates with the \ac{MBS} according to the max-\ac{RSRP} rule, and only inactive \acp{MBS} are shut down. The parameter $\zeta$ is set to be inversely proportional to the number of \acp{UE} in the network. Detailed simulation parameters are provided in Table \ref{simul_params} and are set based on \cite{3GPPTR36.763, 3GPPTR38.811, 3GPPTR38.821,3GPPTR38.901, 3GPPTR36.814, 3GPPTR36.931}.

\begin{table}[h!]
\begin{center}
\begin{tabular}{|l|l|}
\hline Parameter & Value \\
\hline Total Bandwidth $W$  & $40$ $\mathrm{MHz}$ \\
\hline Urban/Rural Inter-Site Distance & $500/1732$ $m$ \\
\hline Number of Macro BSs & $1776$ \\
\hline Satellite Altitude \cite{3GPPTR38.821} & $600$ km \\
\hline Number of arms $n$ & $875$\\
\hline Terrestrial Max Tx Power per RE ${p}_{\mathrm{max}}$ \cite{3GPPTR36.814} & $ 17.7 \text{ } \mathrm{dBm}$ \\
\hline Satellite Max Tx Power per RE ${p}_{\mathrm{max}}$ \cite{3GPPTR38.821} & $15.8 \text{ } \mathrm{dBm}$ \\
\hline Antenna gain (Terrestrial) $G_{\mathrm{T_X}}$ \cite{3GPPTR36.931} & $14\text{ } \mathrm{dBi}$ \\
\hline Antenna gain (Satellite) $G_{\mathrm{T_X}}$ \cite{3GPPTR38.821} & $30\text{ } \mathrm{dBi}$ \\
\hline Shadowing Loss (Terrestrial) $\mathrm{SF}$ \cite{3GPPTR38.901} & $4 \hspace{0.25em} \mbox{--} \hspace{0.25em} 8 ~\mathrm{dB}$ \\
\hline Shadowing Loss (Satellite) $\mathrm{SF}$ \cite{3GPPTR38.811} & $0 \hspace{0.25em} \mbox{--} \hspace{0.25em} 12 ~\mathrm{dB}$ \\
\hline Line-of-Sight Probability (Terrestrial / Satellite) & Refer to \cite{3GPPTR38.901} / \cite{3GPPTR38.811} \\
\hline White Noise Power Density & $-174 \text{ }$ $ \mathrm{dBm} /  \mathrm{Hz}$ \\
\hline Coverage threshold $\mathrm{RSRP}_{\rm min}$ & $-120 \text{ }$ $ \mathrm{dBm}$ \\
\hline Urban/Rural \acp{UE} distribution proportion & $40\% / 60\%$\\
\hline {UE Antenna gain $G_{\mathrm{UE}}$ \cite{3GPPTR38.811}} & {$0\text{ } \mathrm{dBi}$}\\
\hline
\end{tabular}

\end{center}
\caption{Simulation parameters.}
\label{simul_params}
\end{table}

\subsection{UE Satisfaction Analysis}\label{subseq:UE_satisfaction}
\begin{figure}[h]
    \centering
        \includegraphics[width = 0.45 \textwidth]{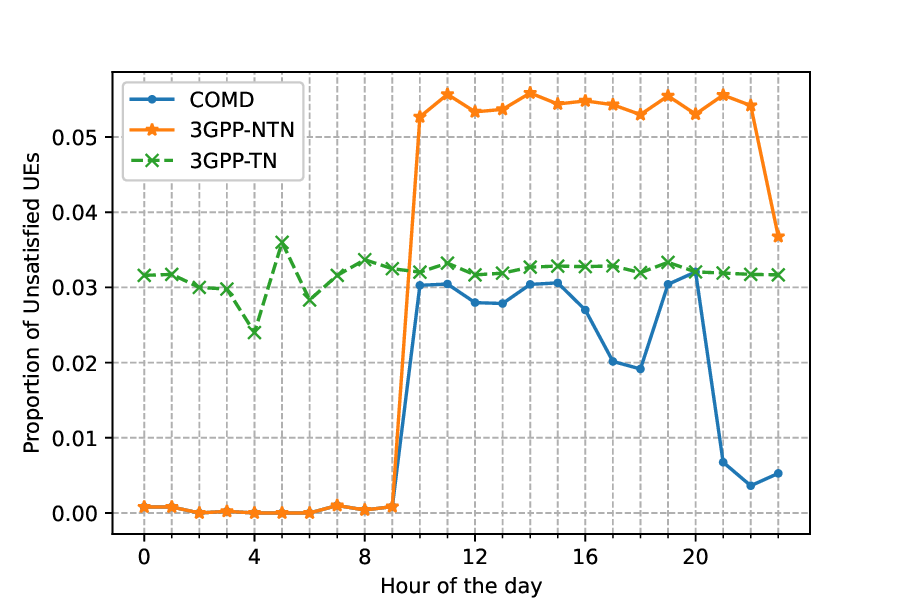}
        \caption{Daily satisfied \ac{UE} proportion profile for various settings.}
        \label{unsat_ues_daily}
         \vspace{-0.2cm}
\end{figure}

Firstly, we study the \ac{UE} satisfaction constraint violation. To that end, Fig. \ref{unsat_ues_daily} depicts the proportion of \acp{UE} who are not satisfied throughout the day for our framework as well as the two benchmarks mentioned previously.
We notice straight away that the proportion of \acp{UE} unsatisfied is steady around $3~\%$ throughout the day for \texttt{3GPP-TN}. This is explained by the fact that this setting does not include a satellite, leading to several cell-edge \acp{UE} being out of coverage and thereby unsatisfied. Conversely, \texttt{3GPP-NTN} and \texttt{COMD} bring this proportion down to nearly $0~\%$ in low-traffic hours ($0$ AM - $9$ AM) solely by the addition of the satellite. However, as the traffic demand increases, we notice that the proportion of unsatisfied \acp{UE} jumps to roughly $6~\%$. Indeed, the max-\ac{RSRP} association does not consider the load of each cell, leading to the overload of the satellite and a deteriorated data-rate for served \acp{UE}. Nevertheless, our framework is able to improve on both benchmarks during high-traffic, as the optimal policy learns the best setting of the parameters $\theta$ ($\alpha$ in particular), leading to a more efficient load distribution and, consequently, a decrease in the number of unsatisfied \acp{UE}.

\subsection{Network Performance Analysis}\label{subseq:Network_Performance}
\begin{figure}[h]
    \centering
        \includegraphics[width = 0.45 \textwidth]{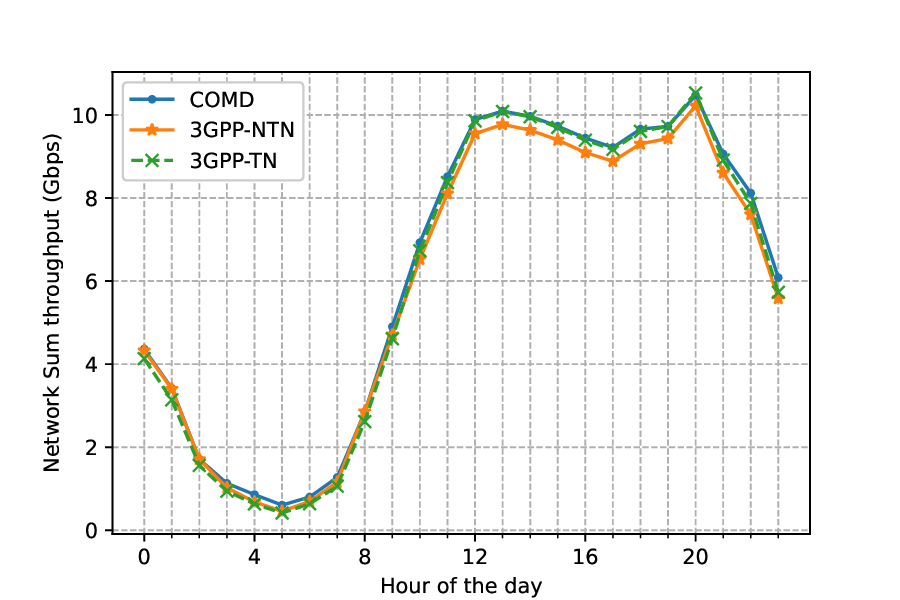}
        \caption{Daily network \ac{ST} profile for various settings.}
        \label{ST_plot}
         \vspace{-0.2cm}
\end{figure}

\begin{figure}[h]
    \centering
        \includegraphics[width = 0.45 \textwidth]{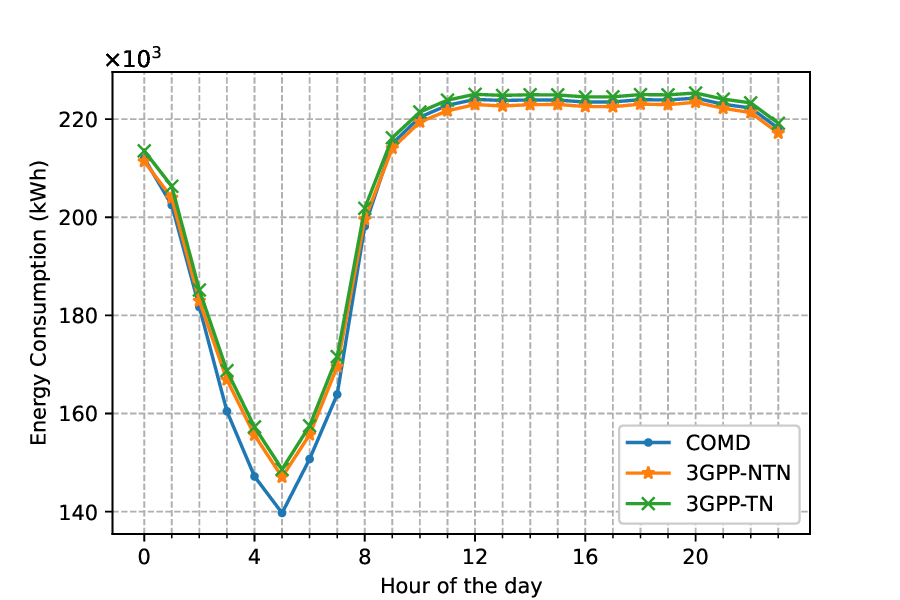}
        \caption{Daily \ac{TN} energy consumption profile for various settings.}
        \label{Energ_conso_plot}
         \vspace{-0.2cm}
\end{figure}

In this section, we analyse the network performance in terms of total achieved capacity, as well as the total \ac{TN} energy consumption. To that end, Fig. \ref{ST_plot} shows the evolution of the \ac{ST} throughout the day, while Fig. \ref{Energ_conso_plot} shows the \ac{TN} energy consumption. 
Since each \ac{UE} has a specific demand, the \ac{ST} is inherently bounded, and we cannot exceed this threshold, which limits the potential gains that can be observed.
The gains are directly explained by the number of unsatisfied \acp{UE} in the network, as we see an average \ac{ST} improvement in high-traffic hours of $1~\%$ and $4~\%$ compared to \texttt{3GPP-TN} and \texttt{3GPP-NTN}, respectively. In low-traffic, the average \ac{ST} gain soars up to $19~\%$ and $10~\%$, respectively. \newline
In terms of energy consumption, we see an average decrease of roughly $5~\%$ compared to both benchmarks in low traffic hours. Again, this is due to our policy learning the best configuration of $\theta$ (especially $\varepsilon$ and $\tau_\nu$) that could facilitate the shutdown of terrestrial \acp{MBS}. 
In high-traffic scenarios, the energy consumption for \texttt{COMD} is slightly worse, as the emphasis is on load balancing. Indeed, the optimal policy selects a $\theta$ configuration which enables handovers to inactive \acp{MBS}, resulting in fewer \acp{MBS} shutdowns compared to the two benchmarks while maintaining a higher satisfaction rate, as seen in Section \ref{subseq:UE_satisfaction}.



\section{Conclusion}
\label{sec:conclusions}
In this work, we proposed a novel framework for online optimisation in integrated \ac{TN-NTN}, aimed at jointly improving \ac{UE} satisfaction through load balancing and reducing \ac{TN} energy consumption.
By formulating the problem as a \ac{MAB} and leveraging the \ac{BCOMD} algorithm, our framework adaptively optimises a set of control parameters to select the most suitable system configuration in response to time-varying network conditions, striking a balance between enhanced capacity and energy efficiency.
Through extensive simulations over a $24$-hour period, we demonstrated that our approach significantly improves performance compared to standard \texttt{3GPP-TN} and \texttt{3GPP-NTN} benchmarks. Notably, our method reduces the proportion of unsatisfied users during peak hours, enables up to $19 ~\%$ higher \ac{ST} and $5 ~\%$ lower energy consumption in low-traffic scenarios.
Our future works will include a theoretical study on the dynamic regret bound that our algorithm can achieve if we change our mirror map to the Tsallis entropy, as well as exploring alternate, more robust estimators for the gradients.


\bibliographystyle{IEEEtran}
\bibliography{reference.bib}

\input{acronyms}

\end{document}

%% file: acronyms.tex
\begin{acronym}[Acronym]
    \acro{EE}{energy efficiency}
    \acro{HetNet}{heterogeneous network}
    \acro{BCOMD}{Bandit-feedback Constrained Online Mirror Descent}
    \acro{DL}{downlink}
    \acro{UL}{uplink}
    \acro{CDF}{cumulative distribution function}
    \acro{NR}{new radio}
    \acro{PL}{path loss}
    \acro{ICC}{international conference on communications}
    \acro{UAV}{unmanned aerial vehicle}
    \acro{UE}{user equipment}
    \acro{PRB}{physical resource block}
    \acro{RE}{resource element}
    \acro{RSRP}{reference signal received power}
    \acro{SINR}{signal-to-interference-plus-noise ratio}
    \acro{QoS}{Quality of Service}
    \acro{UE}{user equipment}
    \acro{LoS}{line-of-sight}
    \acro{NLoS}{non Line-of-sight}
    \acro{3GPP}{3rd Generation Partnership Project}
    \acro{thp}{throughput}
    \acro{UMi}{urban micro}
    \acro{UMa}{urban macro}
    \acro{RMa}{rural macro}
    \acro{CAGR}{compound annual growth rate}
    \acro{HO}{handover}
    \acro{MNO}{mobile network operator}
    \acro{NOMA}{non-orthogonal multiple access}
    \acro{ML}{machine learning}
    \acro{RAN}{radio access network}
    \acro{BS}{base station}
    \acro{MBS}{macro base station}
    \acro{ISD}{inter-site distance}
    \acro{LEO}{low-earth orbit}
    \acro{O2I}{outdoor-to-indoor}
    \acro{MCPA}{multi-carrier power amplifier}
    \acro{RF}{radio frequency}
    \acro{SLT}{sum log-throughput}
    \acro{TN-NTN}{terrestrial and non-terrestrial network}
    \acro{TN}{terrestrial network}
    \acro{NTN}{non-terrestrial network}
    \acro{BCGA}{block coordinate gradient ascent}
    \acro{ST}{sum throughput}
    \acro{HAPS}{high-altitude platform station}
    \acro{MIMO}{multiple-input multiple-output}
    \acro{MAB}{multi-armed bandit}
\end{acronym}